%% file: foreground-subtraction-intensity.tex
\documentclass{PoS}
\usepackage[authoryear]{natbib}
\bibpunct{(}{)}{;}{a}{}{,}
\usepackage{xspace}
\usepackage{subfigure}
\newcommand{\KLfull}{Karhunen-Lo\`{e}ve\xspace}

\let\ifpdf\relax
\newcommand{\tcm}{21\,cm\xspace}

\newcommand{\trans}{{\rm T}}
\newcommand{\bmath}{\mathbf}

\usepackage{savesym}

\usepackage{amsmath}
\savesymbol{tablenum}

\input{mathdef}

\title{Foreground Subtraction in Intensity Mapping with the SKA}

\ShortTitle{Foreground Subtraction in Intensity Mapping}

\author{\speaker{Laura Wolz}$^{1,2}$, Filipe B. Abdalla$^{1,3}$, David Alonso$^2$,
Chris Blake$^4$, Philip Bull$^5$, Tzu-Ching Chang$^6$, Pedro G. Ferreira$^2$,
Cheng-Yu Kuo$^6$, Marios G. Santos$^{7,8}$ and Richard Shaw$^{9}$\\
$^1$Department of Physics and Astronomy, University College London,
London WC1E
6BT, UK\\
$^2$ Astrophysics, University of Oxford, DWB, Keble Road, Oxford OX1 3RH,
UK\\
$^3$ Department of Physics and Electronics, Rhodes University, PO Box 94, Grahamstown, 6140 South Africa\\
$^4$ Centre for Astrophysics \& Supercomputing, Swinburne University of
Technology, P.O. Box 218, Hawthorn, VIC 3122, Australia\\
$^5$ Institute of Theoretical Astrophysics, University of Oslo, P.O. Box 1029
Blindern, N-0315 Oslo, Norway\\
$^6$ Academia Sinica Institute of Astronomy and Astrophysics, P.O. Box 23-141,
Taipei, 10617 Taiwan \\
$^7$ Department of Physics, University of Western Cape, Cape Town 7535, South Africa\\
$^8$ SKA SA, 4rd Floor, The Park, Park Road, Pinelands, 7405, South Africa\\
$^{9}$ Canadian Institute for Theoretical Astrophysics, 60 St. George Street,
Toronto, ON M5S 3H8, Canada\\
\\
E-mail: \email{lwolz@star.ucl.ac.uk}

}

\abstract{\tcm  intensity mapping experiments aim to observe the diffuse neutral hydrogen (HI) distribution on large scales which traces the Cosmic structure. The Square Kilometre Array (SKA) will have the capacity to measure the \tcm signal over a large fraction of the sky. However, the redshifted  \tcm signal in the respective frequencies is faint compared to the Galactic foregrounds produced by synchrotron and free-free electron emission. In this article, we review selected foreground subtraction methods suggested to effectively separate the \tcm signal from the foregrounds with intensity mapping simulations or data. We simulate an intensity mapping experiment feasible with SKA phase 1 including extragalactic and Galactic foregrounds. We give an example of the residuals of the foreground subtraction with a independent component analysis and show that the angular power spectrum is recovered within the statistical errors on most scales. Additionally, the scale of the Baryon Acoustic Oscillations is shown to be unaffected by foreground subtraction.}

\FullConference{
Advancing Astrophysics with the Square Kilometre Array\\
June 8-13, 2014\\
Giardini Naxos, Italy}

\begin{document}
\section{Introduction}
\noindent{Intensity mapping of neutral hydrogen (HI) is a very promising, efficient tool to measure the large-scale structure of the Universe \citep{2004MNRAS.355.1339B, 2009atnf.prop.2491V,2011ApJ...741...70L, Peterson:2012hb}. The HI gas with spectral emission of $\lambda=21\rm cm$ caused by the spin-flip of the valence electron is assumed to linearly trace the Dark Matter distribution. In intensity mapping, the entire HI flux is measured in large resolution elements which still allow observations of structure on the Baryon Acoustic Oscillation scale (BAO) \citep{Wyithe:2007rq, Chang:2007xk,2014arXiv1405.1452B}. These experiments are optimally conducted by telescopes with a large field-of-view (FoV) through, for instance, multi-pixel feeds. The Green Bank telescope (GBT) has pioneered the detection of the large scale structure with the \tcm signal \citep{Chang:2010jp,2013ApJ...763L..20M,2013MNRAS.434L..46S}.  Near future experiments include single dish telescope such as BINGO \citep{2012arXiv1209.1041B} and interferometric designs such as CHIME \citep{Bandura2014} or Tianlai \citep{Chen:2012xu}. The SKA will be able to carry out various intensity mapping experiments at different redshifts either in auto-correlation mode, i.e. single dish observations, or interferometric mode. For a more detailed discussion of intensity mapping with the SKA, we refer to the chapters \cite{SKA:IM} and \cite{SKA:BAOIM}.}

The drawback of intensity mapping observations are the foregrounds mainly due to the high radio emission of
our own Galaxy caused by synchrotron and free-free electron radiation.  In the considered
frequency range, the Galactic foregrounds significantly dominate the acquired
maps by several orders of magnitude. 
The only full-sky observation of the Galactic radio emission is at frequency
$\nu=408\rm MHz$ \citep{Haslam:1982zz}. This map combined with smaller surveys is the basis of most studies to understand and fully model the Galactic radio emission, i.e. \cite{deOliveiraCosta:2008pb,Jelic:2008jg,2014arXiv1405.1751A}.
The varying spectral index with latitude and the lack of small-scale observations impede the comprehension of the Galactic foregrounds. 
The missing observational constraints on the Galactic foregrounds make the
subtraction a crucial, highly sensitive step in the intensity mapping analysis pipeline.
Foreground residuals can cause serious systematic effects which bias the cosmological analysis and
therefore require careful consideration \citep{2014MNRAS.441.3271W}. Most statistical methods employed for this task are based on the spectral smoothness of the
Galactic foregrounds whereas the cosmological signal is expected to have low correlations between frequency bins. Extragalactic foregrounds caused by bright point source emission are expected to show different spatial structure due to gravitational clustering and be far less dominant than Galactic sources. Extragalactic contaminations are yet to be fully incorporated in realistic simulations, however, not considered to cause any significant systematic effects.

In this article, we review three different foreground subtraction approaches: the \KLfull transform (KL, see  \citealt{Shaw2013,2014arXiv1401.2095S}), the singular value decomposition (SVD, see \citealt{2013ApJ...763L..20M}) and the independent component analysis (\textsc{fastica}, see \citealt{2014MNRAS.441.3271W}). The SVD and  \textsc{fastica} are blind search methods which decompose the data in principal or independent components, respectively, whereas the KL transform separates foregrounds and \tcm signal by modeling their statistical properties.  For the KL transform, we reconsider the foreground residuals of the power spectrum of an earlier study. We give an example of foreground removal by applying the \textsc{fastica} to an SKA phase 1 simulation for auto-correlation mode where we chose the settings for SKA1-MID band 1. We emphasize that an SKA1-SUR band 2 experiment would have very similar prospects in terms of foreground removal and results are transferable to such an experiment.
For SKA2 the foreground subtraction for experiments in the low redshift regime $z\approx 1$ will be less challenging due to higher signal-to-noise ratio. Intensity mapping of structure in higher redshifts $z\approx 3$, (which is of particular interest), is expected to be similarly challenging due to the increased sensitivity of the SKA2 experiment.
We present the expected level of foreground residuals and BAO scale measurements in the angular power spectrum  for \textsc{fastica}. In this work, we focus on methods which have been recently applied to intensity mapping simulations or data. In \cite{SKA:FGEOR}, foreground subtraction methods are presented in the context of SKA experiments on the epoch of reionization which potentially can be transferred to the intensity mapping framework.

The article is structured as follows. In Sec.~\ref{sec1}, we describe the \tcm signal, Galactic foregrounds and telescope noise properties including the simulation details. We also review the effect of  instrumental errors on the foreground subtraction. In Sec.~\ref{sec3}, the removal techniques are established and in the following Sec.~\ref{sec4}, the performance of the \textsc{fastica} in foreground subtraction is evaluated by the considering the systematic errors. The conclusions are presented in Sec.~\ref{sec5}.
\section{Description of the Signal Components}\label{sec1}
\noindent In this section, we describe the nature of the cosmological signal we expect to
observe and how we simulate it. We briefly outline the
expected instrumental noise levels and review possible instrumental errors. 
\subsection{21cm Signal}\label{sec21cm}
\noindent The intensity in a frequency bin $\delta\nu$ coming from the 21cm emission of an
object at redshift $z$ with neutral hydrogen mass $M_{\rm HI}$, subtending a solid
angle $\delta\Omega$ is given by \cite{Abdalla:2004ah}
\begin{equation}\label{eq:M2T}
  I(\nu,\hat{\bf n})=\frac{3\,h_p\,A_{12}}{16\pi\,m_{\rm H}}\frac{1}{((1+z)\,r(z))^2}
          \frac{M_{\rm HI}}{\delta\nu\,\delta\Omega}\nu_{21},
\end{equation}
where $A_{12}$ is the Einstein coefficient corresponding
to the emission from the \tcm hyperfine transition, $h_p$ is Planck's constant and
$m_{\rm H}$ is the hydrogen atom mass. Here, $r(z)$ is the
comoving curvature distance $r(z)=c\,{\rm sinn}(H_0\,\sqrt{|\Omega_k|}\,
\chi(z)/c)/(H_0\,\sqrt{|\Omega_k|})$ and $\chi(z)$ is the radial comoving distance
$\label{eq:dzrel}  \chi(z)=\int_0^z{dz'}/{H(z')}$.

This intensity $I(\nu,\hat{\bf n})$ can be written in terms of a black-body
temperature in the Rayleigh-Jeans approximation $T=I\,c^2/(2\,k_B\nu^2)$, where $k_B$
is Boltzmann's constant. Using this we can estimate the mean brightness temperature
coming from redshift $z$ and its fluctuations in terms of the neutral hydrogen density:
\begin{equation}\label{eq:t2d}
  T_{21}(z,\hat{\bf n}) = (0.19055\, {\rm K})\,
  \frac{\Omega_{\rm b}\,h\,(1+z)^2\,x_{\rm HI}(z)}{\sqrt{\Omega_{\rm M}(1+z)^3+
  \Omega_{\Lambda}}}\,(1+\delta_{\rm HI}).
\end{equation}
Here $x_{\rm HI}(z)$ is the neutral hydrogen fraction (i.e. fraction of the total baryon
density in HI) and $\delta_{\rm HI}$ is the HI overdensity field in redshift space
(smoothed over the volume defined by $\delta\nu$ and $\delta\Omega$).

We simulated the cosmological signal by generating a three-dimensional realization of
the neutral hydrogen density in the lightcone using a lognormal field as described in
section 3 of \cite{2014arXiv1405.1751A}. This was done using generic $\Lambda$CDM cosmological
parameters $(\Omega_M,\Omega_b,\Omega_k,h,w_0,w_a,\sigma_8,n_s)=(0.3,0.049,0,0.67,-1,0,0.8,0.96)$,
and we further assumed a redshift dependence for the neutral hydrogen fraction of
$x_{\rm HI}=0.008\,(1+z)$ and a clustering bias of 1 ($\delta_{\rm HI}=\delta$).
Redshift-space distortions were included using the radial velocity field inferred from
the Gaussian overdensity field used for the lognormal realization. The simulation box used
is large enough to encompass a full-sky volume to redshift 2.5 (with the observer placed
at the centre), and has a spatial resolution of about $2.65\,{\rm Mpc}/h$.

We interpolated the density field into spherical temperature maps at different 
frequencies using Eq. \ref{eq:t2d}. These maps were generated using the HEALPix package
\citep{Gorski:2004by} with a resolution parameter ${\tt n\_side}=512$
($\delta\theta\sim0.11^o$), at frequency intervals of $\delta\nu=0.7\,{\rm MHz}$ between
$405\,{\rm MHz}\,\,(z\sim2.5)$ and $945\,{\rm MHz}\,\,(z\sim0.5)$, making up a total of
770 frequency bands.

\subsection{Galactic Foregrounds}
\noindent For the simulation used in this analysis we generated foreground realizations for
Galactic synchrotron emission and free-free emission (both Galactic and extragalactic).
We used the method described in section 4 of \cite{2014arXiv1405.1751A} and did not include
any leakage of the polarized synchrotron radiation.

The free-free foregrounds were generated as a Gaussian random realization of the
corresponding power spectra modelled in \cite{Santos:2004ju}. This models these
foregrounds as isotropic processes, which is obviously not a good approximation
for the Galactic case. However, due to their exceptionally smooth frequency
dependence and subdominant amplitude, we do not believe a more sophisticated
modelling is required at this stage.

Galactic synchrotron is by far the largest foreground for intensity mapping, and
a more complex method was used to simulate it. 
The method is largely based on that of \cite{2014arXiv1401.2095S}, and is also
similar to those used in other studies \citep{2010MNRAS.409.1647J,Shaw2013}. The method starts by extrapolating the 408 MHz Haslam map
\citep{Haslam:1982zz} to other frequencies using a given model for the
direction-dependent synchrotron spectral index (for which we used the
Planck Sky Model, see \citealt{2013A&A...553A..96D}).
Smaller-scale structure and frequency-decorrelation is then added on top of this
through a Gaussian realization of the power spectra in \cite{Santos:2004ju}.
Further details can be found in section 4.2 of \cite{2014arXiv1405.1751A}.
\subsection{Instrumental Noise} % Phil
\noindent The noise RMS (flux sensitivity) for a single-pointing observation with a 
single-dish radio telescope is given by
\begin{equation}
\sigma_S = \frac{2 k_B T_\mathrm{sys}}{A_e \sqrt{\delta \nu t_p}}, \label{eqn:sigmaS}
\end{equation}
where $A_e$ is the effective collecting area of the dish, $t_p$ is the duration of the observation, and we have assumed 
that the noise is Gaussian and uncorrelated. The total system temperature is 
$T_\mathrm{sys} = T_\mathrm{inst} + T_\mathrm{sky}$, where 
$T_\mathrm{inst} \sim \mathrm{few} \times 10 \mathrm{K}$ 
depends on the noise characteristics of the receiver system, and
$T_\mathrm{sky} \approx 60 \mathrm{K} (\nu / 300 \,\mathrm{MHz})^{-2.55}$ 
accounts for the temperature of the sky due to background radio emission. 
Converting Eq. (\ref{eqn:sigmaS}) into a brightness temperature sensitivity 
gives $\sigma_T = T_\mathrm{sys} / \sqrt{\delta \nu t_p}$ in the Rayleigh-Jeans
limit.

In our simulations, we assume that the various SKA configurations will be 
able to perform 10,000 hour auto-correlation surveys over an area of 30,000 sq.
deg\footnote{We focus on auto-correlation surveys here because of their better 
sensitivity to relevant cosmological scales for the SKA Phase 1 configurations; 
see \cite{SKA:IM} for a discussion of the relative merits of auto-correlation 
and interferometric surveys.}. For an instrument with $N_\mathrm{dish}$ 
independent dishes, the total survey area $\Omega_\mathrm{tot}$ can be covered 
with an observation time per pointing of 
$t_p = t_\mathrm{tot} (N_\mathrm{dish} \Omega_B / \Omega_\mathrm{tot})$, where
$\Omega_B \approx \lambda^2 / A_e$ is the beam solid angle of a single dish. 
Assuming no overlap between survey pointings, the brightness temperature 
sensitivity per pointing becomes
\begin{equation}
\sigma_T = \frac{T_\mathrm{sys}}{\sqrt{\delta \nu t_\mathrm{tot}}} \sqrt{\frac{\Omega_\mathrm{tot}}{N_\mathrm{dish} \Omega_B}}.
\label{equ:noise}
\end{equation}
The noise RMS per pixel in our simulations is then given by 
$\sigma_\mathrm{px} = \sigma_T \sqrt{\Omega_B \delta \nu / \Omega_\mathrm{pix} \delta \nu_\mathrm{pix}}$.

We simulate an intensity mapping experiment which can be performed by an SKA1-MID survey with band 1. We assume $T_\mathrm{inst}=28\mathrm{K}$ for $N_\mathrm{dish}=190$ with a dish diameter of $D_\mathrm{dish} =15 \rm m$.
\subsection{Instrumental Effects}\label{sec2}
\noindent In theory, the spectrally smooth foregrounds can be  very well extracted by presented methods, however, instrumental effects significantly complicate the subtraction. 
Most of the foreground subtraction methods are not capable of dealing with varying  beams which is required for auto-correlation observations. The issue can be mitigated by deconvolving the data to the lowest resolution, as has been done for our simulation and for example in \cite{2014MNRAS.441.3271W}. Scale-dependent foreground removal methods are required for future experiments in order to not lose spatial information.

The most prominent instrumental systematics are polarization leakage and frequency-dependent beam distortions which cause spatial and spectral modes to mix. In \cite{2014arXiv1405.1751A}, a first step is taken to include polarization leakage in the intensity mapping simulation such that future studies can test the mode mixing effects.  Additionally,  calibration errors, telescope pointing errors and RFI adulterate signal processing.  Existing methods need to be advanced to include these instrumental effects in the foreground removal step. 
\section{Review of Foreground Removal Methods}\label{sec3}
\noindent In the following section, we review selected foreground removal methods, the SVD, KL transform and \textsc{fastica}. Each of these methods have been designed and applied to different types of intensity mapping data types and are part of power spectrum estimator pipelines.  The KL transform is part of a detailed interferometer simulation for CHIME \citep{Shaw2013,2014arXiv1401.2095S}.  The SVD is a more empirical approach utilized to subtract the foregrounds of GBT observations \citep{2013ApJ...763L..20M, 2013MNRAS.434L..46S}. The \textsc{fastica} has been tested on SKA-like simulations in \cite{2014MNRAS.441.3271W}. Each of these methods work in a different mathematical and experimental framework and we refer the reader to the respective articles for more details on the background description.
\subsection{\KLfull Transform}
\noindent The \KLfull transform has a long history in Cosmology
\citep{1995PhRvL..74.4369B,1996ApJ...465...34V}, but was first suggested as a
method for \tcm foreground removal in \cite{Shaw2013}. It has also been used
for the analogous problem of E/B mode separation for polarisation of the CMB
\citep{2002PhRvD..65b3505L,2003PhRvD..67b3501B}. In this section, we give an
overview of the technique, and discuss  its practical implementation in
\cite{Shaw2013,2014arXiv1401.2095S}.

Ultimately foreground cleaning is simply a matter of finding a subset of our
data within which there is significantly more \tcm signal than there are
astrophysical foregrounds. However, in the presence of mode mixing there is no
immediately apparent representation which separates these signals. The \KLfull
transform (often called the Signal-to-Noise eigendecomposition), gives an
automated way of deriving this basis from the two-point statistics of each
component.

This method requires models for the two-point statistics of both the signal
and the foregrounds on the sky. We denote the matrix representation of the
signal power spectrum as $\mC_{21}$, whereas the foreground model, which
includes both the synchrotron emission from our galaxy, and the contribution
from a background of extragalactic point sources, is written as $\mC_{f}$.
Appropriate models are described in \cite{2014arXiv1401.2095S}

The \KLfull transform seeks to find a linear transformation of the data $\vd'
=\mP \vd$ such that the covariance matrices of the \tcm signal $\mS = \mB
\mC_{21} \mB^\hconj$ and foregrounds $\mF = \mB \mC_f \mB^\hconj$ are jointly
diagonalised. That is
\begin{equation}
\mS \rightarrow \mS' = \mP \mS \mP^\hconj = \mLambda \;,
\end{equation}
and
\begin{equation}
\mF \rightarrow \mF' = \mP \mF \mP^\hconj = \mI \; ,
\end{equation}
where $\mLambda$ is a diagonal matrix, and $\mI$ is the identity. In this
diagonal basis we can simply compare the amount of power expected in each mode
by the ratio of the diagonal elements (this is given by the corresponding
entries of $\mLambda$), and identify the regions of the space with low
foreground contamination (large entries in $\mLambda$).

\begin{figure}

\begin{center}
\includegraphics[width=0.7\textwidth]{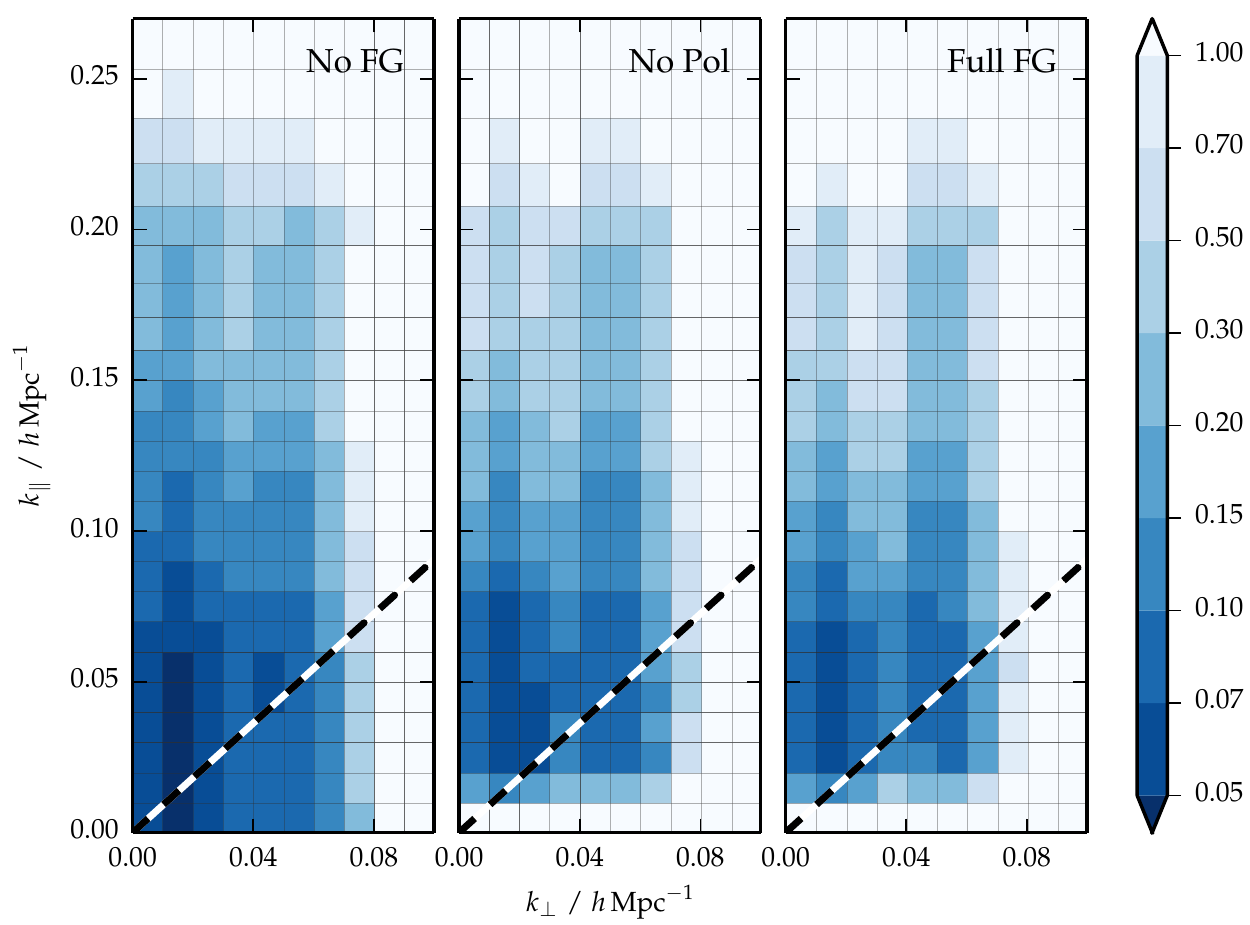}
\end{center}
\caption{Forecast errors on the power spectrum as a fraction of its fiducial
value for a $40\,\mathrm{m} \times 40\,\mathrm{m}$ cylinder telescope similar to \cite{Bandura2014}, observing between $400$--$500\,\mathrm{MHz}$. The three panels show the
predicted errors without foregrounds (left), with unpolarised foregrounds
(centre), and with fully polarised foregrounds (right). We can see that we lose sensitivity to the large scale line of sight mode. The dashed line
indicates the predicted bound of the `foreground wedge', showing that with
perfect knowledge of our instrument, foregrounds can be successfully cleaned
well into this region.}
\label{fig:kl_ps}
\end{figure}

This transformation can be found by solving the generalised eigenvalue problem
$\mS \vx = \lambda \mF \vx$. This gives a set of eigenvectors $\vx$, and
corresponding eigenvalues $\lambda$. Writing the eigenvectors in a matrix
$\mP$, row-wise, gives the transformation matrix to diagonalise the
covariances. The eigenvalues $\lambda$ corresponding to each eigenvector give
the diagonal matrix $\mLambda$. To isolate the \tcm signal, we can simply select modes with eigenvalue
(signal-to-foreground power) greater than some threshold. In Fig.~\ref{fig:kl_ps} we show the effect of foreground cleaning on the power spectrum errors of simulated cylinder telescope similar to \cite{Bandura2014}.

The \KLfull transform is a general and very effective scheme for the component
separation. Unfortunately, the full covariance matrices are dense, and very
large, $O(N_\text{pix} N_\text{freq}) \sim 10^8$--$10^9$ on a side, making the
diagonalisation, which is an $O(N^3)$ operation, impossible in the general
case. However, this technique can still be applied to the restricted domain of
\emph{transit telescopes}, where the symmetries of the system allow the
problem to be broken up into many smaller problems which are significantly
more tractable. This decomposition is known as the $m$-mode formalism, and was
first demonstrated in \cite{Shaw2013}. Overall the complexity is reduced to
$O(N_\text{pix}^2 N_\text{freq}^3)$ saving a factor of $\sim 10^6$ in
computation.
\subsection{Singular Value Decomposition}
\noindent In this section, we present the foreground cleaning formalism in \cite{2013MNRAS.434L..46S} for the redshifted HI survey with the Green Bank Telescope (GBT).  It is based on the idea that the \tcm signal is a line emission and has structure in the frequency/redshift domain, while the foregrounds are smooth on those scales. One can thus separate the two components in a model independent way by making use of principal component analysis. In this case, we arrange the three-dimensional maps into a frequency-frequency covariance matrix and use singular value decomposition to identify the dominant frequency modes. We label them as \lq\lq{foregrounds}\lq\lq{} and project out their contribution to obtain foreground cleaned maps.

Specifically, we rearrange the three-dimensional map into an $N_\nu \times N_\theta$ matrix $\mat{M}$, where $N_\theta$ includes all two-dimensional spatial pixels. For the purpose of this comparison, we ignore thermal noise in the map. The empirical $\nu-\nu’$ covariance of the map is ${\mat{C} =\mat{M} \mat{M}^\trans/ N_\theta}$, which contains both foregrounds and \tcm signal.  Making use of the component separation idea, the matrix can be factored as $\mat{C} = \mat{U} \mLambda \mat{U}^\trans$, where $\mLambda$ is diagonal and contains the eigenvalues in descending order. We tag the first few modes as 'foregrounds',  and from each line of sight, we can then subtract a subset of the modes $\mat{U}$ that describe the largest components of the frequency variance through the operation $(1-\mat{U} \mat{S} \mat{U}^\trans) \mat{M}$, where $\mat{S}$ is a selection matrix with $1$ along the diagonal for modes to be removed and $0$ elsewhere.

In practice, the separation of foreground and signal modes is not perfect due to the imperfect characterization of instrumental response.  The choice of the selection matrix $\mat{S}$ is a compromise between maximal foreground removal and minimal signal loss.  To estimate the latter, we inject simulated \tcm signal to the data stream to determine the transfer function $T$ which describes loss of \tcm signal due to foreground removal.  As a rule of thumb, $T =P_{\rm sig.\,out} / P_{\rm sig.\,in} \sim [(1 - N_{\rm m}/N_\nu)(1 - N_{\rm m}/N_{\rm res})]^2$, where $N_{\rm m}$ is the number of modes removed, $N_\nu$ is the number of frequency channels and $N_{\rm res}$ is the number of angular resolution elements. A limited number of resolution elements can greatly reduce the efficacy of the foreground cleaning at the expense of signal.  The details of foreground transfer function calculation and the signal compensation can be found in \cite{2013MNRAS.434L..46S}.

Our approach to foreground removal is limited by the amount of information in the maps. The fundamental limitation here is not simply from the number of degrees of freedom along the line of sight, but instead is limited by the smaller of independent angular or frequency resolution elements in the map \citep{2009ASPC..407..389N}. To see why this is the case, notice that in the absence of noise, our cleaning algorithm is equivalent to taking the SVD of the map directly: ${\mat{M} = \mat{U} \msigma \mat{V}^\trans}$ and thus $\mat{C} \propto \mat{M} \mat{M}^\trans = \mat{U} \msigma^2 \mat{U}^\trans$, with the same set of frequency modes $\mat{U}$ appearing in both decompositions.  The rank of $\mat{C}$ coincides with the rank of $\mat{M}$ and is limited by the number of either angular or frequency degrees of freedom.
%%
%%
%%%%
%%%
\subsection{Independent Component Analysis}
\noindent In the following subsection, the basic principles of the fast independent component analysis \textsc{fastica}
method  \citep{DBLP:journals/tnn/Hyvarinen99} are outlined. \textsc{fastica} has been successfully used for foreground removal in the astrophysical framework, see for example \cite{Maino:2001vz, Bottino:2009uc, Chapman:2012yj}.

\textsc{fastica} is a method designed to decompose mixed signals into their independent constituents. Statistical independence is measured by the use of the Central Limit theorem which states that the probability distribution density (pdf) of a sum of independent variables is always more Gaussian than the pdf of one single component. The inverse application of that theorem implies that a single component can be extracted from the mixture by maximizing the non-Gaussianity of the pdf. This is turn implies that \textsc{fastica} can not be applied to Gaussian variables.
We can formulate the problem in the linear equation $\bmath x= \mathbf A \bmath s$, where $\mathbf x$ is the signal, $ \mathbf A$ the mixing matrix and $\bmath s$ the independent components (IC). \textsc{fastica} blindly solves the inverted problem 
\begin{equation}
\bmath s= \mathbf W\bmath x
\label{eq:invica}
\end{equation}
where the weighing matrix $\mathbf W$ defined as the inverse of $\mathbf A$ is unknown. As a measure of the non-Gaussianity of the ICs, \textsc{fastica}  maximizes a proxy for the negentropy. 

\textsc{fastica} has been employed to subtract foregrounds in intensity mapping simulations in \cite{2014MNRAS.441.3271W}. Since the foregrounds are highly correlated between frequencies, \textsc{fastica} incorporates them into the ICs. The \tcm signal has a very low  correlation which is close to Gaussian. Hence, \textsc{fastica} blindly reconstructs the foregrounds and the residuals of this analysis are cosmological signal plus the receiver noise. In  \cite{2014MNRAS.441.3271W}, it has been shown that foreground subtraction with  \textsc{fastica} does not affect cosmological distance measures such as the BAO scale. An residual analysis of the SKA simulations with \textsc{fastica} is demonstrated in the following sections.
\section{Residual Errors and BAO recovery}\label{sec4}
\noindent In this section, we present the foreground residuals and systematic errors when applying \textsc{fastica} to the described SKA1-MID simulation.
\subsection{Experimental Set-up}
\noindent The SKA1-MID band 1 simulation is based on the \tcm and Galactic foregrounds simulation presented in \cite{2014arXiv1405.1751A}, where the frequency range is $406{\rm MHz}<\nu<945\rm MHz$ with a frequency width of $\delta\nu=0.7\rm MHz$ per channel. The beam is approximated by a Gaussian beam with constant solid angle $\Omega_{\rm beam}=1.6\mathrm{deg}^2$. The instrumental noise is simulated according to Equ.~\ref{equ:noise} for an SKA-MID band 1 like observation. We mask out the sky above $63\rm deg$ latitude in equatorial coordinates which results in a coverage of $30,000\mathrm{deg}^2$. The intensity maps are stacked into constant redshift bins with width $\delta z \approx 0.05$ after the foreground removal to increase the signal-to-noise in the systematics analysis.
\subsection{Results}
\begin{figure}
\subfigure[Relative error on $C(\ell)$ for $1.45<z<1.50$]{
\includegraphics[width=0.5\textwidth, trim=20 20 20 50]{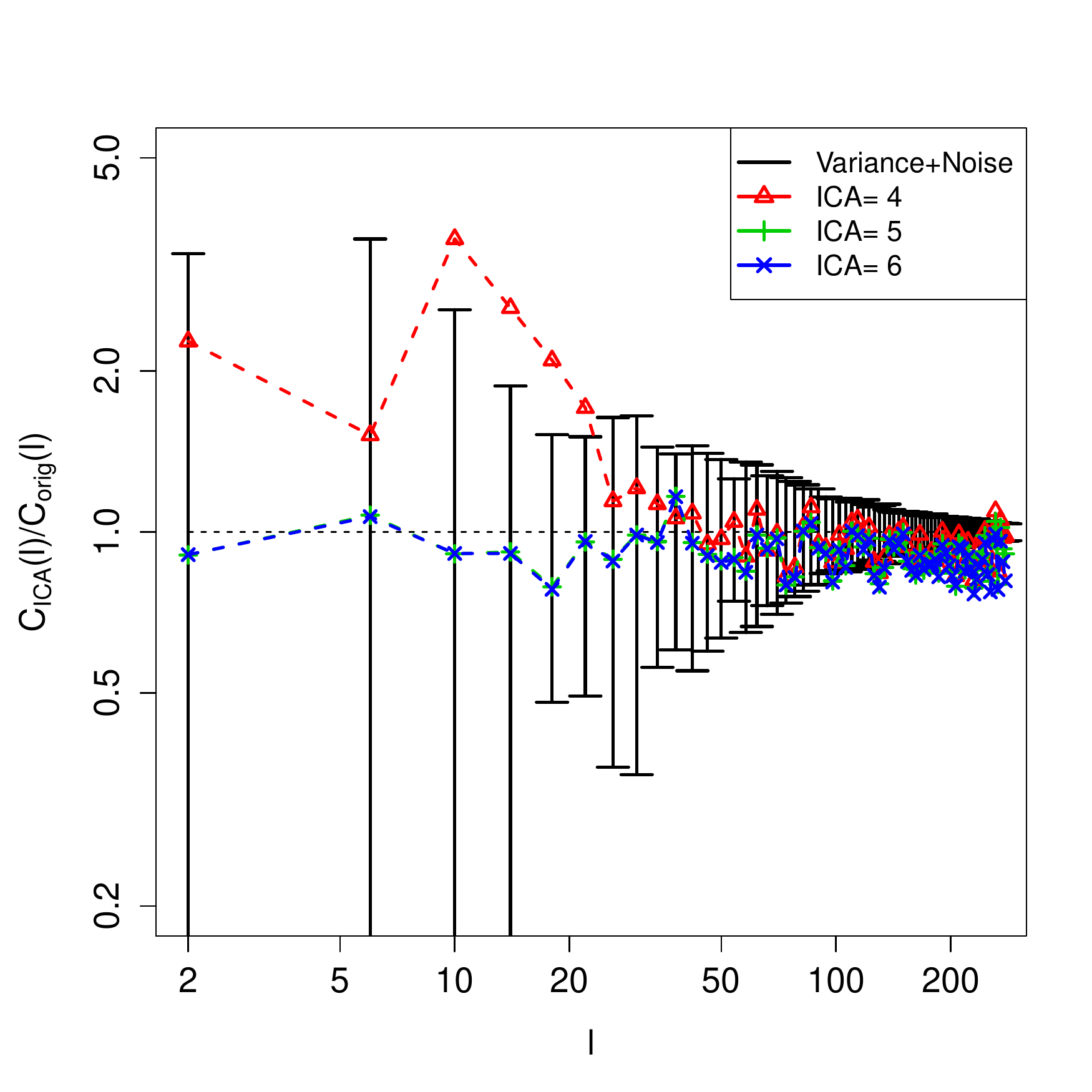}
\label{fig:clfastica}
}\subfigure[Relative systematic errors; \textsc{fastica} with IC=5]{
\includegraphics[width=0.5\textwidth, trim=20 20 20 50]{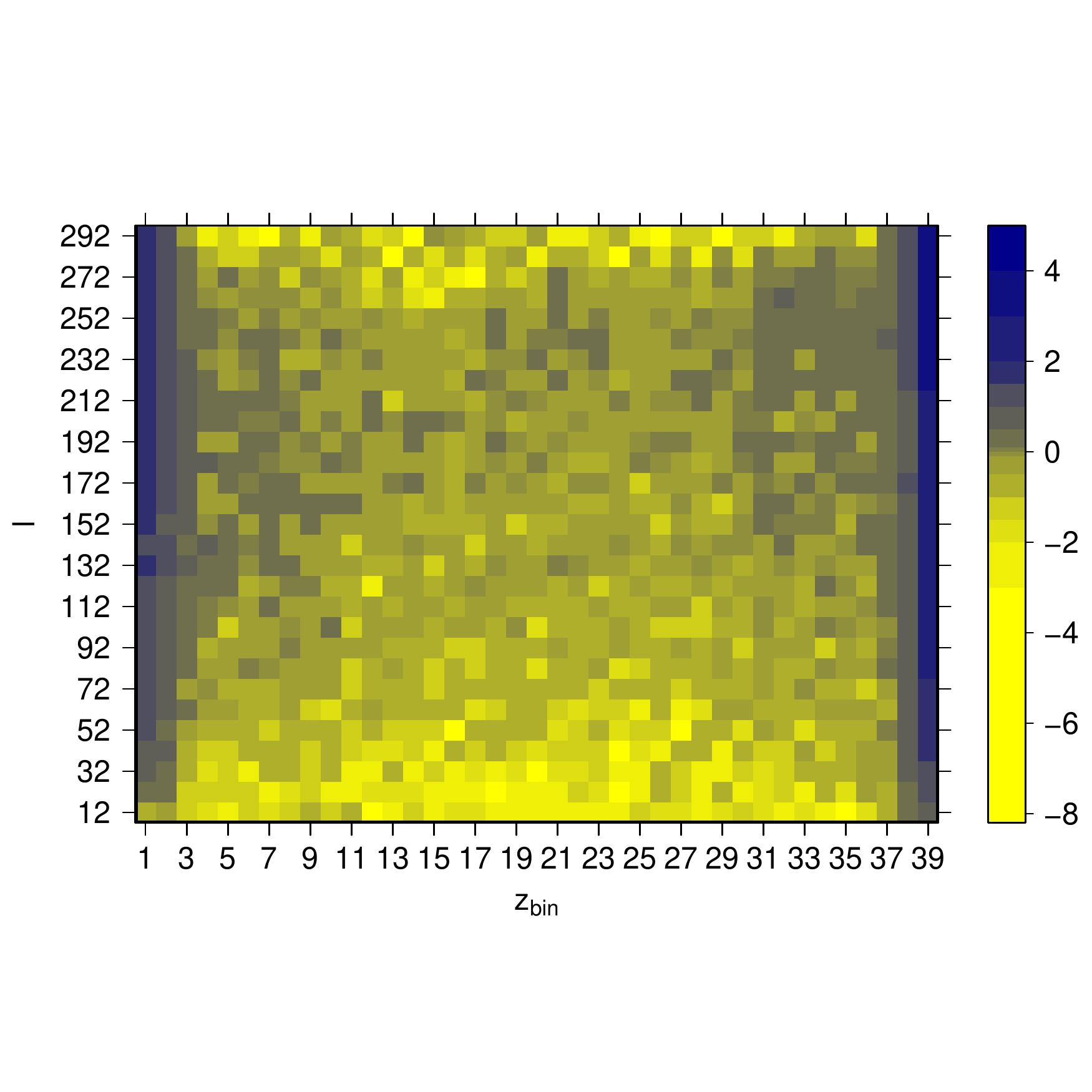}
\label{fig:sysfastica}
}
\caption{The relative errors of the foreground subtraction with \textsc{fastica} on the angular power spectrum as a function of scale. On the left side, we present the fraction $C_{\rm ICA}(\ell)/C_{\rm orig}(\ell)$ for different numbers of ICA for an example redshift bin. The relative systematic errors $\delta(\ell, z_i)$ are given on a natural logarithmic scales in the right panel.}
\label{fig:cl}
\end{figure}

\noindent We evaluate the foreground subtraction by estimating the angular power spectrum $C(\ell)$ of the original \tcm intensity maps and the reconstructed maps. We correct for the partial sky coverage with the Peebles approximation, as described in \cite{2014MNRAS.441.3271W}. In Fig.~\ref{fig:clfastica}, we show the relative error $C_{\rm ICA}(\ell)/C_{\rm orig}(\ell)$ of the \textsc{fastica}-cleaned maps for different number of ICs. The black error bars indicate the statistical variance of the measurements given by sampling error and instrumental noise. It can be seen that for a number of ICs higher than 4 the reconstruction converges towards the original input $C(\ell)$ for scales $\ell>100$. For smaller scales, the broadband spectrum is slightly distorted and we observe deviations of $C_{\rm ICA}(\ell)$ outside the statistical errors. 

In Fig.~\ref{fig:sysfastica}, the relative systematic error $\delta(\ell, z_i)$ is depicted for all redshift bins $z_i$ on the $x$-axis and for multipole bins $\delta \ell=20$ in the $y$-direction. The systematic error is defined as $\Delta_{\rm{sys}}=|C_{\rm{orig}}(\ell)-C_{\rm{ICA}}(\ell)|$. We divide that error by the statistical uncertainty on the measurements $\sqrt{\sigma(C(\ell))}$ composed of sampling variance and instrumental noise. The relative errors are pictured in a logarithmic scale such that values larger than zero imply insignificant systematic errors and smaller than zero indicate scales where systematics dominate the statistical errors. We see that the foregrounds at the edges of the redshift bins are poorly subtracted due to incomplete frequency information and those regions should be neglected in subsequent analysis.

\begin{figure}
\subfigure[BAO scale in angular power spectrum; \textsc{fastica} with IC=5]{
\includegraphics[width=0.5\textwidth, clip=true, trim=50 30 43 50]{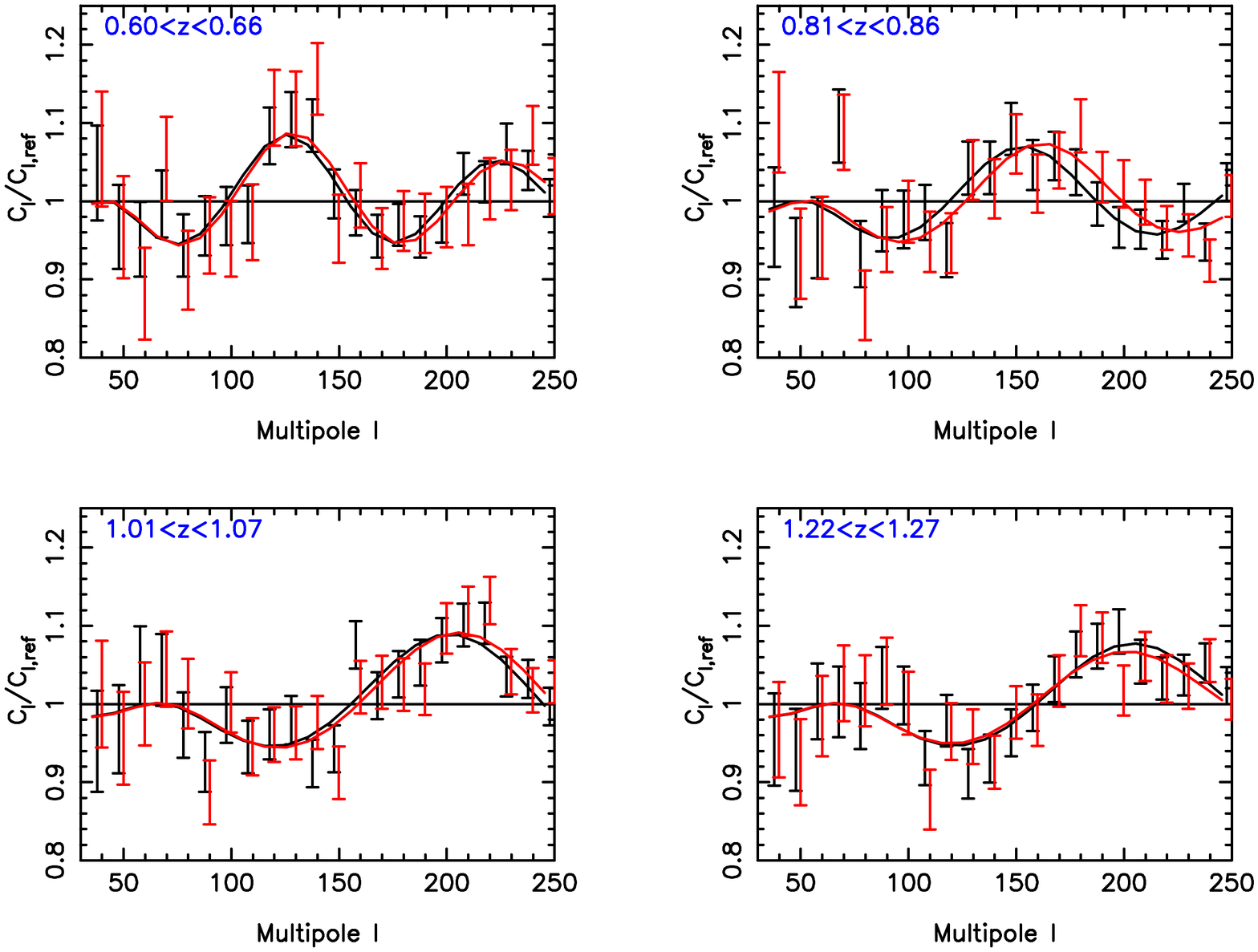}
\label{fig:baowiggles}
}\subfigure[BAO fit; \textsc{fastica} with IC=5]{
\includegraphics[width=0.5\textwidth, clip=true, trim=20 20 20 5]{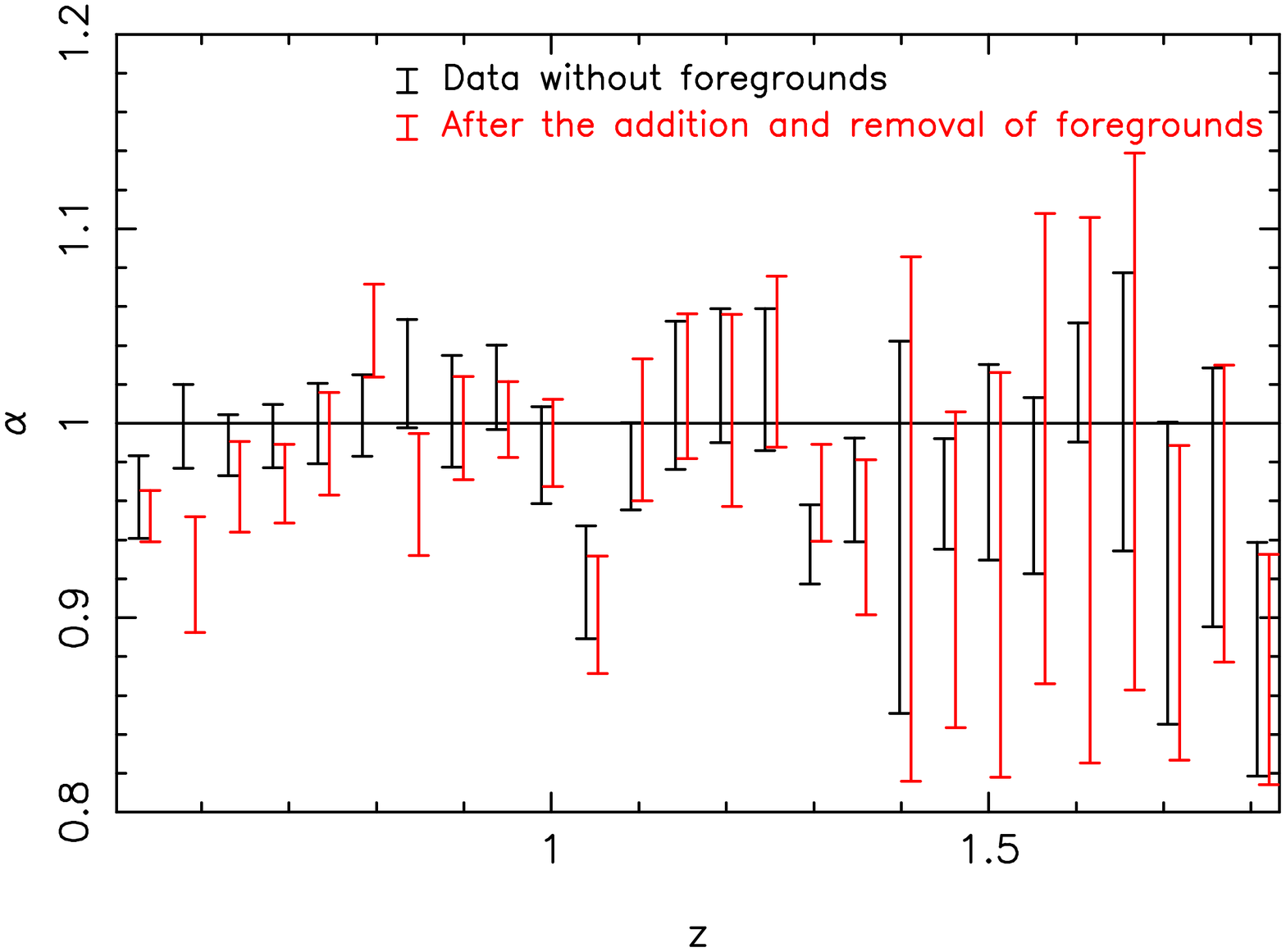}
\label{fig:baofit}
}
\caption{The BAO measurements of the original simulations without foregrounds (black) are compared to the measurements including foregrounds (red). In the left panel, we present the BAO wiggles in the angular power spectrum divided by a featureless fiducial model. The lines are produced by fitting a model to the measurements. The right panel shows the BAO scale distortion parameter $\alpha$ without and with foregrounds as a function of redshift.  }
\label{fig:bao}
\end{figure}

Furthermore, we present BAO scale measurements of the SKA1 simulation in Fig.~\ref{fig:bao}, the details of the methods can be found in \cite{2014MNRAS.441.3271W}. In Fig.~\ref{fig:baowiggles}, the BAO wiggles in the power spectrum produced by dividing by a BAO-featureless power spectrum model are shown for 4 different redshift bins. The lines represent the fitted models to the estimated data points. The black measurements mark the simulation without any foreground contaminations and the red lines include systematic errors. We see that there is no significant shift of the BAO features induced by the foreground subtraction. Fig.~\ref{fig:baofit} presents the scale distortion parameter $\alpha$ as a function of redshift where $\alpha=D_A(z_i)/D_{A,\rm fid}(z_i)$. The BAO recovery is not significantly affected by the foreground subtraction for most redshifts $z_i$.

\section{Conclusions}\label{sec5}
\noindent In this article, we have presented a review of three effective methods for subtracting foregrounds in intensity mapping experiments. We have simulated a future SKA1 intensity mapping experiment with special focus on realistic foreground modeling. We presented a detailed residual analysis of the foreground-removed \tcm signal with \textsc{fastica}.  We conclude the article as follows.
\begin{itemize}
\item \textbf{Method comparison: }
The KL transform is based on a-priori knowledge on the properties of foregrounds and \tcm signal. SVD and \textsc{fastica} blindly decompose the observed data such that foregrounds are removed by subtracting an empirically chosen number of principal/independent modes. The blind methods are excellent in dealing with unexplored foregrounds or foreground fluctuations, however, prone to remove \tcm signal or leave foreground residuals in the data. By detailed, realistic simulations and residual analysis such as presented in this work, the errors can be significantly reduced and systematics well-understood. The KL transform has the advantage of separating the components based on their covariances which reduces described errors. This method is less versatile to experimental settings due to its high computational costs and can possibly be affected by poor foreground modeling.
\item \textbf{Systematic errors: }
All presented methods have been shown to effectively clean intensity maps from high Galactic foregrounds in either simulations or observations such as the SVD method on the GBT data. Figs.~\ref{fig:kl_ps} and \ref{fig:cl} show that the systematic errors mostly affect the very large scales perpendicular to the line-of-sight. KL transform and \textsc{fastica} have been shown to not significantly bias the BAO detection in the cleaned data. We emphasize the importance of comprehensive and detailed simulations of intensity mapping experiments to ensure the reliability of the cosmological analysis.
\item \textbf{Future requirements: }
Instrumental errors can influence the data analysis significantly by mode-mixing effects. The foreground removal can be corrupted by telescope errors such as polarization leakage, telescope pointing and calibration errors. \cite{2014arXiv1401.2095S} has included detailed instrumental effects in their simulation pipeline and studied the influence of each error on the power spectrum. Such effects require careful simulations for each individual experimental set-up in order to understand the systematic effects and how they interact with the foreground removal.
\end{itemize}
In this article, we outlined some of the latest efforts to subtract foregrounds from intensity mapping data. We showed that the \tcm power spectrum can be successfully recovered via a wide range of scales and the BAO scale measurements are not biased by foreground subtraction.  Within the next decade, those methods need to be advanced to the requirements of SKA1 and SKA2 experiments with special attention to include polarization effects and instrumental errors into the systematic analysis.
\section*{Acknowledgments}
\noindent LW is supported by the IMPACT fund. FBA acknowledges the support of the Royal
Society via a University Research Fellowship. DA is supported by European Research Council grant 259505. CB acknowledges the support of the
Australian Research Council through the award of a Future Fellowship. PB is supported by European Research Council grant StG2010-257080. PGF acknowledges
support from the Higgs centre, STFC, BIPAC and the Oxford Martin School. MGS acknowledges support by the South African Square Kilometre Array
Project, the South African National Research Foundation and FCT grant
PTDC/FIS-AST/2194/2012.
\bibliographystyle{apj}
\bibliography{bib}
\end{document}

%% file: mathdef.tex
\usepackage{amsmath}
\usepackage{bm}
\usepackage{mathtools}

\renewcommand{\vec}[1]{\mathbf{#1}}
\newcommand{\mat}[1]{\mathbf{#1}}
\newcommand{\matg}[1]{\bm{#1}}

\newcommand{\hconj}{\ensuremath{\dagger}}

\newcommand{\vx}{\vec{x}}

\newcommand{\vd}{\vec{d}}

\newcommand{\mS}{\mat{S}}

\newcommand{\mC}{\mat{C}}
\newcommand{\mB}{\mat{B}}

\newcommand{\mF}{\mat{F}}
\newcommand{\mI}{\mat{I}}

\newcommand{\mP}{\mat{P}}

\newcommand{\msigma}{\matg{\Sigma}}
\newcommand{\mLambda}{\matg{\Lambda}}